\documentclass[runningheads]{llncs}

	\usepackage{fancyvrb}
	\usepackage{color}

	\usepackage{graphicx}
	\usepackage{amssymb, amsmath}
	\usepackage{enumerate}
	\usepackage[noend]{algpseudocode}
	\usepackage{algorithm, refcount}
	 \usepackage[colorlinks]{hyperref}
	\usepackage{xspace}
	\usepackage{tabularx}
	
	\algdef{SE}[DOWHILE]{Do}{doWhile}{\algorithmicdo}[1]{\algorithmicwhile\ #1}
	 
	\newcommand {\R}   {\mathbb R}
	
	\newcommand {\C}   {\mathbb C} 
	\newcommand {\Z}   {\mathbb Z}
	\newcommand {\N}   {\mathbb N}
	\newcommand {\G}   {\mathbb G}

	\newcommand {\ii}  {{\mathbf i}}

	\newcommand{\contDisc}[1]{\Delta(#1)}

	\newcommand   {\pol}     {P}
	\newcommand   {\sep}[1]  {\textnormal{sep}(#1)}

	\renewcommand {\deg}     {d}
	\newcommand   {\var}     {z}
	\renewcommand {\root}    {\alpha}
	\newcommand   {\rootrad} {r}
	 
	\newcommand   {\ann}     {A}
	\newcommand   {\annset}     {\mathcal{A}}
	
	\newcommand {\Graeffe}[2] {{#1}^{[#2]}}
    
    \newcommand {\shift}[2] { {#1}{#2} }
	
	\newcommand{\ass}{\leftarrow}
	\newcommand{\app}[1]{\widetilde{#1}}

	\newcommand{\Tstart}{T^*}
    \newcommand{\Tzerot}{T^0}
	\newcommand{\CCzerot}{C_\C^0}
	
	\newcommand{\CRzerot}{C_\R^0}

	\newcommand{\Tstar}[1]{T^*(#1)}
	\newcommand{\Tzero}[1]{T^0(#1)}
	\newcommand{\CCzero}[1]{C_\C^0(#1)}
	
	\newcommand{\CRzero}[1]{C_\R^0(#1)}
	\newcommand{\CRstar}[1]{C_\R^*(#1)}

	\newcommand{\RRC}{RRC~}
	\newcommand{\RRCstar}{RRC$^*$~}
	\newcommand{\solveRRC}{{\tt SolveRRC}}
	\newcommand{\solveRRCstar}{{\tt SolveRRC^*}}

	\newcommand{\RI}[1]{{\color{magenta}#1}}
	\renewcommand{\RI}[1]{{#1}}

	\newcommand{\cored}[1]{{\color{red}#1}}
	\newcommand{\coblue}[1]{{\color{blue}#1}}
	\newcommand{\coma}[1]{{\color{magenta}#1}}
	
	\newcommand{\ignore}[1]{}
	
	\newcommand{\ccluster}{\texttt{Ccluster}\xspace}
	
	\newcommand{\cclusterR}{\texttt{CclusterR}\xspace}
	\newcommand{\cclusterRR}{\texttt{CclusterR}\xspace}

	\newcommand{\clang}{\texttt{C}\xspace}
	\newcommand{\mpsolve}{\texttt{MPSolve}\xspace}
	\newcommand{\anewdsc}{\texttt{ANewDsc}\xspace}
	\newcommand{\machine}{\texttt{Intel(R) Core(TM) i7-8700 CPU @ 3.20GHz}\xspace}
	
	\newcommand{\risolate}{\texttt{Risolate}\xspace}
	\newcommand{\risolateRR}{\texttt{RisolateR}\xspace}

\usepackage{breqn}
\newtheorem{Definition}{Definition}
\newtheorem{Remark}[Definition]{Remark}
\newtheorem{Theorem}[Definition]{Theorem}
\newtheorem{Corollary}[Definition]{Corollary}
\newtheorem{Proposition}[Definition]{Proposition}

\renewcommand{\qed}{\hfill$\Box$}
\usepackage{lineno}
\let\oldequation\equation
\let\oldendequation\endequation

\renewenvironment{equation}
  {\linenomathNonumbers\oldequation}
  {\oldendequation\endlinenomath}
  
\renewenvironment{equation*}
  {\linenomathNonumbers\oldequation}
  {\oldendequation\endlinenomath}
  
\begin{document}
 
\title{
Root Radii and Subdivision 
for Polynomial Root-Finding
\thanks{Victor's work is supported by NSF Grants CCF 1563942 and CCF 1733834 and  PSC CUNY Award 63677 00 51}
}

\authorrunning{R. Imbach and V. Pan}
\author{R\'emi Imbach\inst{1}
  \and Victor Y. Pan\inst{2}
     \institute{Courant Institute of Mathematical Sciences, New York
                University, USA\\
                \email{remi.imbach@nyu.edu},
                \and 
                Lehman College and Graduate Center of City University of New York\\
                \email{victor.pan@lehman.cuny.edu} }
                }
\maketitle

\begin{abstract}
We depart from our approximation of 2000 of all root radii of a polynomial, which has readily extended Sch{\"o}nhage's  efficient algorithm of 1982 for a single root radius.
We revisit this extension, advance it, based on our simple but novel idea, and yield significant practical acceleration of the known near optimal subdivision  algorithms for complex and real root-finding of user's choice. We achieve this by means of significant saving 
of exclusion tests  and Taylor's shifts, which are the bottleneck of subdivision root-finders. This saving  relies on our novel recipes for the initialization of root-finding iterations of independent interest. We demonstrate our practical progress with numerical tests, provide extensive analysis of the resulting algorithms, and show that, like the preceding  subdivision  root-finders, they support near optimal Boolean complexity bounds. 

\keywords{ 
Real root isolation, 
Complex root clustering,
Root radii algorithm,
Subdivision iterations
}

\end{abstract}


\section{Introduction}
\label{sec_intro}

\subsubsection*{Overview.} The recent subdivision iterations for univariate polynomial Complex Root Clustering (CRC) and Real Root Isolation (RRI) approximate all roots in a fixed Region of Interest (RoI) and, like the algorithm of Pan (1995, 2002), achieve near optimal bit complexity for the so called benchmark problem. Furthermore
they allow robust implementations, one of which is currently the user's choice for solving the RRI problem, including the task of the approximation of all real roots.
\RI{Another implementation, for the CRC problem,
is slower  by several orders of magnitude than the package MPSolve (the user's choice)
for the task of finding all complex roots.
However it outperforms MPSolve for solving the CRC problem where the RoI contains only a small number of
roots.}
We significantly accelerate these highly efficient root-finding iterations by applying
 our novel techniques for their initialization.
Next we specify the background and outline our contributions.

\subsubsection*{Polynomial roots and root radii.}

For  a polynomial $\pol$
of degree $\deg$ in 
$\Z[\var]$
and a Gaussian integer 
$c\in\G\RI{:=\{a+\ii b ~|~ a\in\Z, b\in\Z\}}$,
 let $\root_1(\pol,c),\ldots,\root_{\deg}(\pol,c)$ be the 
$\deg$ non-necessarily distinct roots
of $\pol$
such that
\begin{equation*}
 |\root_1(\pol,c)-c|\geq |\root_2(\pol,c)-c| 
 \geq \ldots \geq 
 |\root_{\deg-1}(\pol,c)-c|\geq |\root_{\deg}(\pol,c)-c|.
\end{equation*}
For all $1\leq i\leq \deg$, write $\rootrad_i(\pol,c):=
|\root_i(\pol,c)-c|$, 
$\root_i(\pol):=\root_i(\pol,0)$
and $\rootrad_i(\pol):=\rootrad_i(\pol,0)$,
so that
\begin{equation*}
 \rootrad_1(\pol,c)\geq \rootrad_2(\pol,c) 
 \geq \ldots \geq 
 \rootrad_{\deg-1}(\pol,c)\geq \rootrad_{\deg}(\pol,c).
\end{equation*}
Then
\begin{center} \fbox{
		\begin{minipage}{0.95 \columnwidth} \noindent
		\textbf{Root Radii Covering (RRC) Problem}\\ \noindent
		\textbf{Given:} 
                        a polynomial 
                        $\pol\in\Z[z]$ of degree $\deg$,
                        a Gaussian integer 
                        $c\in\G$,
		                a real number $\delta>0$\\
		\noindent \textbf{Output:} 
		$d$ positive
		numbers $\rho_{c,1},\ldots,\rho_{c,\deg}$ satisfying
		\begin{equation}
		\label{eq_outputRRC}
		 \forall s=1,\ldots,\deg,~~\frac{\rho_{c,s}}{1+\delta} \leq \rootrad_s(\pol,c) 
		                       \leq (1+\delta)\rho_{c,s}.
		\end{equation}
	\end{minipage}
} \end{center} 
$\rho_{c,1},\ldots,\rho_{c,d}$
of Eq.~(\ref{eq_outputRRC}) for fixed
$c\in\G$ and $\delta>0$
define $\deg$ 
possibly overlaping
concentric annuli. 
The connected components of their union
form
a set $\annset_c$ of 
$\deg_c\leq\deg$ disjoint concentric annuli
centered at $c$.
They 
cover all roots of $\pol$ and are
said to be an \emph{annuli cover}
of the roots of $\pol$. We
are going to  use them
in subdivision root-finding iterations.

\subsubsection*{Two Root-Finding Problems.}
We
count roots 
with  multiplicity and
consider discs 
$D(c,r):=\{z ~|~ |z-c|\leq r\}$
on the complex plane.
For  a positive $\delta$
  let $\delta\Delta$ and $\delta B$ denote
the concentric
$\delta$-dilation of a disc $\Delta$
and a 
real line segment (\emph{i.e.} interval) $B$. 
Then
\begin{center} \fbox{
		\begin{minipage}{0.95 \columnwidth} \noindent
		\textbf{Complex Root Clustering (CRC) Problem}\\ \noindent
		\textbf{Given:} a polynomial $\pol\in\Z[z]$ of degree $\deg$,
		                $\varepsilon>0$\\
		\noindent \textbf{Output:} $\ell\leq \deg$ couples
		$(\Delta^1,m^1),\ldots,(\Delta^\ell,m^\ell)$ satisfying:
		\begin{itemize}
		 \item the $\Delta^j$'s are pairwise disjoint
		discs of radii $\leq \varepsilon$,
		 \item $\Delta^j$ and $3\Delta^j$ contain $m^j>0$ roots,
		 \item each complex root of $\pol$ is in a $\Delta^j$.
		\end{itemize}
	\end{minipage}
	} \end{center}
 
\begin{center} \fbox{
		\begin{minipage}{0.95 \columnwidth} \noindent
		\textbf{Real Root Isolation (RRI) Problem}\\ \noindent
		\textbf{Given:} a polynomial $\pol\in\Z[z]$ of degree $\deg$\\
		\noindent \textbf{Output:} 
		$\ell\leq \deg$ couples $(B^1,m^1),\ldots,(B^\ell,m^\ell)$
        satisfying:
        \begin{itemize}
         \item the $B^i$'s are disjoint real line segments,
         \item each $B^i$ contains a unique real root of multiplicity $m^j$,
         \item each real root of $\pol$ is in a $B^i$.
        \end{itemize}
	\end{minipage}
} \end{center}
It is quite common 
to  pre-process $\pol\in\Z[\var]$ in order  to make it square-free, with $m^j=1$ for all $j$, but we do not use this option. 
We can state both CRC and RRI problems  for $\pol$ with rational coefficients and readily reduce them to the above versions with integer coefficients by scaling.

Write $\|\pol \|:=\|\pol\|_\infty$
and call the value $\log_2\|\pol \|$ the
\emph{bit-size} of $\pol $.

\subsubsection*{The benchmark problem.}
For the bit-complexity of  the so called \textit{benchmark} root-finding problem of the isolation  of all roots of a square-free $\pol\in\Z[\var]$
of degree $\deg$ and 
bit-size $\tau$
 the record  bound of 1995 (\cite{pan2002univariate}) is $\app{O}(\deg^2(\deg+\tau))$, near optimal for $\tau>\deg$ and
  based on a divide and conquer approach. It was reached again in 2016 
(\cite{becker2018near,becker2016complexity}),
 based on subdivision iterations
 and
 implemented in \cite{ICMSpaper}.
 
 \subsubsection*{Our contributions}
 We first present and 
 analyze an algorithm
 \RI{\solveRRC~}
 that solves the RRC problem 
for polynomials with integer coefficients and any fixed center $c\in\G$. Our algorithm is adapted
from 
\cite{pan2000approximating} (which  
has extended  Sch\"onhage's 
highly efficient approximation
 of a single root radius in \cite{schonhage1982fundamental}) 
to simultaneous approximation of  all
 $\deg$ root radii. Our specialization of this root radii algorithm 
to the case of
 integer polynomials and 
our analysis of its  bit-complexity
are novel.
\RI{
We use \solveRRC~ for $\delta\in\deg^{-O(1)}$
and $|c|\in O(1)$; under such assumptions,
it solves the RRC problem with a bit-complexity
in $\app{O}(\deg^2(\deg + \tau))$.
}

We then 
 improve 
solvers for the RRI and the CRC problems
based on subdivision
with annuli covers that we compute
 by applying \solveRRC.
 The complexity of
subdivision root-finders is dominated at 
its bottleneck stages of  root-counting and particularly exclusion tests, at which costly {\em Taylor's shifts}, aka the shifts of the variable, are applied. We significantly accelerate the 
root-finders  for both RRI and CRC 
problems by means of 
using 
fewer exclusion
 tests and calls for root-counting and hence fewer Taylor's shifts.
We achieve this by limiting complex root-finding to the intersection of three  annuli covers of the roots centered in $0, 1$ and $\ii$
and by limiting
real root-finding to the intersection of a single annuli cover
centered in $0$
with the real line.

\begin{table}[t!]
\centering
 \begin{tabular}{rrr||cc||ccc||cc||ccc||}
       &     &          & \multicolumn{2}{c||}{\risolate}  
                        & \multicolumn{3}{c||}{\risolateRR}
                        & \multicolumn{2}{c||}{\ccluster}  
                        & \multicolumn{3}{c||}{\cclusterRR}\\\hline
$\deg$ & $\tau$ & $d_\R$   & $t$ & $n$
                        & $t$ & $n$ & $t'$
                        & $t$ & $n$ 
                        & $t$ & $n$ & $t'$ \\\hline
\multicolumn{13}{c}{Bernoulli polynomial} \\\hline
512 & ~2590 & ~124 & ~6.15 & ~672 
                   & ~0.38 & ~17    & ~0.25 
                   & ~136  & ~13940 
                   & ~50.7 & ~4922  & ~7.59\\\hline
\multicolumn{13}{c}{Mignotte polynomial} \\\hline
512 & 256  & 4   & 1.57 & 49    
                 & 1.67 & 14 & 0.27
                 & 88.8 & 10112
                 & 28.3 & 2680   & 3.05\\\hline
 \end{tabular}
 \caption{Runs of \risolate, \risolateRR, \ccluster and \cclusterRR on two polynomials. \ccluster and \cclusterRR are called with input $\varepsilon=2^{-53}$.
 }
\label{table_intro}
\vspace*{-1cm}
\end{table}

Our improvements are implemented within the 
\clang library \ccluster\footnote{\url{https://github.com/rimbach/Ccluster}}
which provides an eponymous solver for the CRC problem
and a solver for the RRI problem called 
\risolate.
Our novel solvers
are called 
below \cclusterRR and \risolateRR,
and
in Tab.~\ref{table_intro} we overview
how those two solvers perform
 against \ccluster and \risolate
on a Bernoulli and a Mignotte polynomial.
For each test polynomial we show its 
degree $\deg$, its bit-size $\tau$, and the number $\deg_\R$ of real roots.
For each solver, $t$ denotes 
the sequential running time in seconds on 
an \machine machine with Linux and $n$ denotes
the total number of 
Taylor's shift required in the subdivision process.
For \cclusterRR and \risolateRR,
$t'$ is the time spent on
solving the RRC problem with \solveRRC.

We compute the annuli covers 
in a pre-processing step 
by applying algorithm \solveRRC~ for input relative width $\delta = 1/\deg^2$.
This choice of 
$\delta$ is empiric,
and in this sense our improvement of subdivision is heuristic.
From a theoretical point of view,
this allows our algorithms for solving the RRI
and the CRC problems
to support a near optimal bit-complexity.
From a practical point of view, this allows 
us
to significantly reduce the running time
of solvers
based on subdivision by using fewer
Taylor's
shifts in exclusion and root-counting tests,
as we 
highlighted in Table~\ref{table_intro}
(see the columns $t$ and $n$).

The distance  between  roots
of a polynomial of degree $\deg$ and bit-size $\tau$
can be way less than $1/\deg^2$
(see for instance \cite{Mignotte1995});
thus by computing with \solveRRC~ intervals
that  contain
the root radii
of relative width $\delta = 1/\deg^2$,
we do not intend to separate the roots of input polynomials,
and our improvement has no effect 
in the cases
where 
distances between  some roots are less
than $\delta$.
We illustrate this in Table~\ref{table_intro}
for a Mignotte polynomial that has four real roots 
among which two roots have a pairwise distance that is close 
to the theoretical separation bound.
Most of the computational effort in a subdivision solver for 
real roots isolation is spent
on the separation of the
close roots,
and this remains true 
where we use
annuli covers
with relative width larger than the roots separation.

We compare our implementation
with \anewdsc (see \cite{Kobel}, implementing \cite{sagraloff2016computing}) 
and \mpsolve (see \cite{bini2014solving}, implementing Ehrlich's iterations),
which are the current user's choices for solving 
the RRI and the CRC problems, respectively.


\subsubsection*{Related work}
\label{subsec_state}
We departed from the subdivision 
polynomial root-finding for the CRC and RRI 
problems in \cite{becker2016complexity} and \cite{sagraloff2016computing}, resp., and from the algorithms for the \RRC problem in \cite{schonhage1982fundamental} (see \cite{schonhage1982fundamental}[Cor.~14.3],
and \cite{gourdon:inria-00074820}[Algo.~2]) and \cite{pan2000approximating}. 
We achieved practical
progress by complementing these advanced works  with our novel techniques for efficient computation of $O(1)$  annuli covers of the roots. 
 We rely on the customary framework 
 for the analysis of root-finding algorithms and cite  throughout the relevant sources  of our auxiliary techniques.
 
\subsubsection*{Organization of the paper.}

In Sec.~\ref{sec_rootradii} we describe an algorithm for solving
the \RRC problem. 
In Secs.~\ref{section_RRI} and ~\ref{section_CRC} we present our  algorithms for solving the RRI and
CRC problem, respectively.
In Subsec.~\ref{subsec_notations} we
introduce  definitions. 
In Subsec. \ref{subsec_subdivision} we briefly describe the subdivision algorithm for the CRC problem
of \cite{becker2016complexity} and its adaptation  to the solution of
the RRI problem.

\subsection{Definitions}
\label{subsec_notations}

\noindent
Write
$
 \pol :=\pol(\var)
   := \pol_\deg \prod_{i=1}^{\deg} (\var - \root_i(\pol))
   = \sum_{j=0}^{\deg} \pol_j \var^j
$.

\subsubsection*{Root squaring iterations.}
For a positive integer $\ell$ write 
\[
 \Graeffe{\pol}{\ell} := 
 (\pol_\deg)^{2^\ell} \prod_{i=1}^{\deg} (\var - \root_{i}(\pol)^{2^\ell})
\]
and so
$\Graeffe{\pol}{0} = \pol$,  
$\Graeffe{\pol}{\ell} = \Graeffe{(\Graeffe{\pol}{\ell-1})}{1}$
for $\ell\geq 1$, and
\[
 |\root_1( \Graeffe{\pol}{\ell} )|\geq |\root_2( \Graeffe{\pol}{\ell} )| 
 \geq \ldots \geq 
 |\root_{\deg-1}( \Graeffe{\pol}{\ell} )|\geq |\root_{\deg}( \Graeffe{\pol}{\ell} )|.
\]
$\Graeffe{\pol}{\ell}$ is called
the $\ell$-th root squaring iteration of $\pol$, aka
the $\ell$-th Dandelin-Lobachevsky-Gr\"affe (DLG) iteration of $\pol$.

Write
$\Graeffe{\pol}{\ell-1}=\sum_{j=0}^{\deg} (\Graeffe{\pol}{\ell-1})_j \var^j$,
$\Graeffe{\pol}{\ell-1}_e=\sum_{j=0}^{\lfloor\frac{\deg}{2}\rfloor} (\Graeffe{\pol}{\ell-1})_{2j} \var^j$
and 
$\Graeffe{\pol}{\ell-1}_o=\sum_{j=0}^{\lfloor\frac{\deg-1}{2}\rfloor} (\Graeffe{\pol}{\ell-1})_{2j+1} \var^j$.
$\Graeffe{\pol}{\ell}$ can be computed iteratively 
based on
the formula:
\begin{equation}
\label{eq_compDLG}
 \Graeffe{\pol}{\ell} = (-1)^d \left[
 (\Graeffe{\pol}{\ell-1}_{e})^2
 - \var (\Graeffe{\pol}{\ell-1}_{o})^2
 \right].
\end{equation} 
The $j$-th coefficient $(\Graeffe{\pol}{\ell})_j$ of $\Graeffe{\pol}{\ell}$ is related to the coefficients
of $\Graeffe{\pol}{\ell-1}$ 
by:
\begin{equation}
 \label{eq_defbk}
 (\Graeffe{\pol}{\ell})_j = 
 (-1)^{\deg-j}(\Graeffe{\pol}{\ell-1})_j^2 + 
 2\sum_{k=\max(0,2j-\deg)}^{j-1} (-1)^{\deg-j}(\Graeffe{\pol}{\ell-1})_k(\Graeffe{\pol}{\ell-1})_{2j-k}
\end{equation}

\subsubsection*{$L$-bit approximations.}

For any number $c\in\C$, we say that $\app{c}\in\C$
is an $L$-bit approximation of $c$ if
$\|\app{c}-c\|\leq 2^{-L}$.
For a polynomial $\pol\in\C[\var]$, 
we say that
$\app{\pol}\in\C$ 
is an $L$-bit approximation of $\pol$
if $\|\app{\pol}-\pol\|\leq 2^{-L}$,
or equivalently if $\|\app{\pol_j}-\pol_j\|\leq 2^{-L}$
for all $j$.

\subsubsection*{Boxes, Quadri-section,  Line segments, Bi-section}
$[a-w/2, a+w/2] + \ii [b-w/2, b+w/2]$ is the box $B$ of width $w$  centered at $c=a+\ii b$.
The  disc $\contDisc{B}:=D(c,\frac{3}{4}w)$ is a \emph{cover} of $B$.

Partition  $B$ into  four congruent  boxes
(children of $B$), of width $w/2$ and centered at 
$(a\pm\frac{w}{4})+\ii (b\pm\frac{w}{4})$.

$\contDisc{B}:=D(c,w/2)$ is 
the minimal disc
that covers a real line segment $B:=[c-w/2, c+w/2]$ of 
width $w$  centered at $c\in\R$.
  
Partition the segment $B$ into  two segments (children of $B$) of width $w/2$ centered at 
$(c\pm\frac{w}{4}))$.

Let $\mathcal{C}$ be a connected component of 
boxes (resp. real line segments); $B_{\mathcal{C}}$
is
the  minimal box (resp. real line segment) covering $\mathcal{C}$.

\subsection{Subdivision approach to root-finding}
\label{subsec_subdivision}

\cite{becker2016complexity} describes an algorithm for solving a local 
version of the CRC problem: 
for an initial RoI $B_0$ (a box)
it finds clusters 
of roots 
with pairwise distance less than $\varepsilon$ in a small inflation of $B_0$.
Since our CRC problem is for input polynomials in $\Z[\var]$,
one can define a RoI containing all the roots by
using, for instance, the Fujiwara bound
(see \cite{fujiwara1916obere}).

\subsubsection*{Subdivision iterations}
The algorithm in \cite{becker2016complexity} uses subdivision iterations or Quad-tree algorithms 
(inherited from 
\cite{henrici1969uniformly}, see also \cite{pan2000approximating}),
which
constructs a tree rooted in the RoI $B_0$
whose nodes are sub-boxes of $B_0$.
A node $B$ is  \emph{included} only if $2B$ contains
a root, \emph{excluded} only if it contains no root.
A node is \emph{active} if
it is neither included
nor excluded.
At the beginning of a subdivision iteration,
each active node $B$ is tested for exclusion.
Then active boxes are grouped into connected components,
and
for each connected component $\mathcal{C}$
such that $4\contDisc{B_{\mathcal{C}}}$ intersect no other 
connected component, 
a root-counter 
is applied to $2\contDisc{B_{\mathcal{C}}}$.
If $2\contDisc{B_{\mathcal{C}}}$ contains $m>0$ roots
and $\contDisc{B_{\mathcal{C}}}$
has radius less than $\varepsilon$, 
then $(\Delta(B_{\mathcal{C}}),m)$ is returned
as a solution and the boxes of $\mathcal{C}$
are marked as included.
Each remaining active node is
quadrisected into its four active children, to which
 a new subdivision iteration is applied.
Incorporation of Newton's iterations enables quadratic convergence toward clusters
of radii $\varepsilon$.

\subsubsection*{Solving the RRI problem}
Using a root separation lower bound (e.g., of \cite{Mignotte1995}), one can derive 
from \cite{becker2016complexity}
a solution of the RRI problem based on the symmetry  
of roots of $\pol\in\Z[\var]$ 
along 
the real axis.
Let disc $D(c,r)$ with $c\in\R$ contain 
$m$ roots of $\pol$.
For $m=1$  the root in $D(c,r)$ is real.
If $m\geq1$ and $r\leq \sep{\pol}$,
then $D(c,r)$ contains a real root of multiplicity $m$,
where $\sep{\pol}$ is a root separation lower bound
for $\pol$.
For the RRI problem, the RoI $B_0$ is a line segment, and the subdivision tree of $B_0$  is built by means of segment bisection.

\subsubsection*{The $\Tzerot$ and $\Tstart$ tests}
In the algorithm of \cite{becker2016complexity}, the exclusion test 
and root counter are based on  
Pellet's theorem (see \cite{becker2018near}).
For a 
disc $\Delta=D(c,r)$, the counting test $\Tstar{\Delta,\pol}$
returns an integer $k\in\{-1,0,\ldots,\deg\}$ such that
$k\geq 0$ only if $\pol$ has $k$ roots in $\Delta$.
A result $k=-1$ accounts for a failure and holds 
when some roots of $\pol$ are close to the boundary of $\Delta$.
For a given disc $\Delta$, the exclusion test $\Tzero{\Delta,\pol}$
returns $0$ if $\Tstar{\Delta,\pol}$ returns $0$
and returns $-1$ if $\Tstar{\Delta,\pol}$ returns a non-zero integer.
The $\Tstart$ of \cite{becker2018near}
takes 
as an input an $L$-bit approximation of $\pol$ 
and 
with working absolute precision $L$ 
performs about $\log\log \deg$
DLG iterations of the Taylor's shift $\pol(c+r\var)$
of $\pol$. Write $L(\Delta,\pol)$ for the precision $L$
required to carry out the $\Tstart$-test.
Based on Pellet's theorem we obtain the
following results.

\begin{Proposition}[see \cite{becker2018near}, Lemmas 4 and 5]
\label{prop_TStarCost}
 Let $B$ be the box (or real line segment) 
  centered in $c$ with width $w$.
 The total cost in bit operations
  for carrying out $\Tstar{\Delta(B),\pol}$ or
 $\Tzero{\Delta(B),\pol}$ is bounded by
 \begin{equation}
 \label{eq_costTstar}
  \app{O}( \deg( \log\|\pol\| + \deg\log\max(1,|c|,w) + L(\Delta,\pol))).
 \end{equation}
 
 $\Tstar{\Delta(B),\pol}$ returns an integer $k\in\{ -1,0,\ldots,\deg\}$; if $k\geq 0$ then $\Delta(B)$ contains $k$ roots;
 if $k=-1$ then $\pol$ has a root in $2B\setminus(1/2)B$.
 $\Tzero{\Delta(B),\pol}$ returns an integer $k\in\{ -1,0\}$; if $k = 0$ then $\pol$ has no root in $\Delta(B)$;
          if $k=-1$ then $\pol$ has a root in $2B$.
\end{Proposition}

\subsubsection*{Bit complexity}  
 Prop.~\ref{prop_TStarCost}  
 enables one to 
bound the Boolean cost of  exclusion tests and root-counting as well as the size of the subdivision tree  and hence 
the cost of
solving the benchmark
problem in 
\cite{becker2016complexity}.

By applying subdivision iterations  with an exclusion test and a root counter
satisfying Prop.~\ref{prop_TStarCost}
one yields an algorithm with the
same bit-complexity as the algorithm of \cite{becker2016complexity}, namely, $\app{O}(\deg^2(\deg+\tau))$
for the benchmark problem.

\subsubsection*{Implementations}
A modified version of \cite{becker2016complexity} for the CRC
problem has been implemented and made
public within the library \ccluster.
An implementation of the modified algorithm of \cite{becker2016complexity} solving the RRI problem,
called \risolate, is also available within \ccluster.

\section{Root radii computation}
\label{sec_rootradii}

We describe and analyse an 
algorithm for solving the \RRC problem
for a $\pol\in\G[\var]$.
Let $c\in\G$ and 
$\shift{\pol}{c}(\var):=\pol(c+\var)$, so that 
$\rootrad_s(\pol,c)=\rootrad_s(\shift{\pol}{c})$ for all $1\leq s\leq d$.
Hence the \RRC problem for a $c\neq0$ reduces
to the \RRC problem for $c=0$ at the cost of shifting the variable. 

The next remark reduces the \RRC problem for 
$c=0$ and any $\delta>0$ to the \RRC problem 
for $1+\delta=4\deg$ by means of DLG iterations:
\begin{Remark}
 \label{rem_numofGraeffeItts_app}
 Let $g=\lceil \log \dfrac{\log(4\deg)}{\log(1+\delta)}\rceil$,
 let $\rho'>0$ such that there exist an $s$ with:
 \[
		 \frac{\rho'}{4\deg} < \rootrad_s(\Graeffe{\shift{\pol}{c}}{g}) 
		                     < (4\deg)\rho'.
 \] 
 Define $\rho=(\rho')^{\frac{1}{2^g}}$ and
 recall that
 $
  \rootrad_s(\Graeffe{\shift{\pol}{c}}{g}) = \rootrad_s(\pol,c)^{2^g}
 $.
 Then
 \[
		 \frac{\rho}{1+\delta} < \rootrad_s(\pol,c) < (1+\delta)\rho.
 \]
$g$ is in $O(\log\deg)$
if $\delta$ is in $\deg^{-O(1)}$ 
 (for instance, $\delta\ge\deg^{-1}$ or $\delta\ge\deg^{-2}$).
\end{Remark}

Now define the \RRCstar problem as the \RRC problem for $1+\delta=4\deg$ and $c=0$:
\begin{center} \fbox{
		\begin{minipage}{0.95 \columnwidth} \noindent
		\textbf{\RRCstar problem}\\ \noindent
		\textbf{Given:} a polynomial $\pol\in\G[z]$ of degree $\deg$, satisfying $\pol(0)\neq 0$\\
		\noindent \textbf{Output:} 
		$\deg$ positive real numbers $\rho_1',\ldots,\rho_\deg'$ satisfying
		\[
		 \forall s=1,\ldots,\deg,~~\frac{\rho_s'}{4\deg} < \rootrad_s(\pol) 
		                     < (4\deg)\rho_s'.
		\]
	\end{minipage}
} \end{center}
In this setting, we assume that 
$0$ is not a root of $\pol$ and thus $\pol_0\neq0$.
When $0$ is a root of multiplicity $m$, then 
$\rootrad_\deg(\pol)=\ldots=\rootrad_{\deg-m+1}(\pol)=0$
and $\pol_0=\ldots=\pol_{m-1}=0$,
which is easily detected (since $\pol\in\G[\var]$)
and treated accordingly.

In Subsec.~\ref{subsec_tasksp}
we 
recall
an algorithm  
$\solveRRCstar$ 
satisfying:
\begin{Proposition}
\label{prop_tasksp} 
Algorithm 
$\solveRRCstar$ in Algo.~\ref{algo:taskSprime}
 solves the \RRCstar problem
by involving
 $O(\deg\log\|\pol\|)$ bit operations.
\end{Proposition}
In Subsec.~\ref{subsec_proof_prop} we 
prove this proposition.
In Subsec.~\ref{subsec_tasks} we 
present an algorithm $\solveRRC$
satisfying:

\begin{Theorem}
 \label{Th_solving_taskS}
 The algorithm  $\solveRRC$ of Subsec.~\ref{subsec_tasks}
 solves the \RRC problem for $\delta = \deg^{-2}$  at a Boolean cost in 
  \[
  \app{O}( \deg^2(\deg\log(|c|+1) + \log\|\pol\| ) ).
  \]
 This bound turns into 
  $
  \app{O}( \deg^2(\deg + \log\|\pol\| ) )
  $ for $|c| \in O(1)$ and 
   into
  $
  \app{O}( \deg^2\log\|\pol\| )
  $ for  $|c|=0$.
\end{Theorem}

Below we will use root radii computation as a pre-processing step for Complex Root Clustering and Real Root Isolation.
For Real Root Isolation, we use $\solveRRC$ to 
compute 
an annuli cover centered 
at $0$.
For Complex Root Clustering, we use $\solveRRC$ to 
compute three
annuli covers 
with the three centers $0, 1, \ii$.
According to our analysis  of  the \RRC problem, the cost of its solution for 
 $O(1)$  centers $c$ such that
  $|c|\in O(1)$
is dominated by  
a near optimal bit-complexity of root-finding.

For $c=0$, our algorithm 
has a larger bit complexity  
than the  algorithm of
\cite{schonhage1982fundamental} (see \cite{schonhage1982fundamental}[Cor.~14.3]
and \cite{gourdon:inria-00074820}[Algo.~2]), which is in $\app{O}(\deg^2\log^2\deg)$
when $\log\|\pol\|\in O(\deg)$.
Our algorithm, however, 
computes $\deg$ root radii at once
where Sch\"onhage's algorithm 
computes only a single
 root radius.
It is not clear 
whether the latter algorithm 
can be extended to an algorithm 
that would
solve 
the 
RRC problem within
the same bit-complexity
bound.
\subsection{Solving the \RRCstar problem}
\label{subsec_tasksp}

Recall that 
$\pol=\sum_{i=0}^{\deg}P_i\var^i$
and define, for $i=0,\ldots,\deg$,
\begin{equation*}
 p_i = \left\{
 \begin{array}{ll}
  \log|\pol_i| & \text{ if } \pol_i\neq 0,\\
  -\infty      & \text{ otherwise.}
 \end{array}
 \right.
\end{equation*}
According to the following result,
adapted from Prop. 4.2 and its proof in \cite{pan2000approximating}, one can solve
the \RRCstar problem 
by computing the upper part
of the convex hull of the set of points 
$\{ (i,p_i) | i=0,\ldots,\deg \}$
and assuming that the
points $(i,-\infty)$ lie below any line in the plane.

\begin{Proposition}
 \label{prop_edgeofCH_int}
 Given an integer $s$, 
 let $t'$ and $h'$ be integers s.t. 
 \begin{enumerate}[$(i')$]
  \item $t'<\deg+1-s\leq t'+h' \leq \deg$, and
  \item $\forall 0\leq i \leq \deg$,
 the point $(i,p_i)$
 lies below the line
 $((t',p_{t'}), (t'+h',p_{t'+h'}))$.
 \end{enumerate}
 Then $\rho_s'=\left|\dfrac{\pol_{t'}}{\pol_{t'+h'}}\right|^{\frac{1}{h'}}$ satisfies:
 $
		 \dfrac{\rho_s'}{2\deg} < \rootrad_s(\pol) 
		                     < (2\deg)\rho_s'
 $.
\end{Proposition}

Call $CH$ the upper part of the convex hull of the points
$\{ (i,p_i) | i=0,\ldots,\deg \}$,
\RI{and remark that for a given integer $s$, the integers $t'$ and $t'+h'$
satisfying $(i')$ and $(ii')$
in Prop.~\ref{prop_edgeofCH_int} 
are the absciss\ae~of the endpoints of the segment of $CH$
above $(s,p_s)$.}
$CH$ can be computed exactly (the 
$\pol_i$'s are Gaussian integers).
However for solving the \RRCstar problem,
it is sufficient to compute
the upper part of the convex hull of
$M$-bit approximations of the $p_i$'s
with $M\geq 1$.
For $i=0,\ldots,\deg$, define
\begin{equation}
\label{eq_appi}
 \app{p_i}: = \left\{
 \begin{array}{ll}
  M\text{-bit approximation of } p_i & \text{ if } |P_i|>1,\\
  0                  & \text{ if } |P_i|=1,\\
  -\infty            & \text{ otherwise}.
 \end{array}
 \right.
\end{equation}
Let $\app{CH}$ be the upper part of the convex hull of 
 $\{ (i,\app{p_i}) | i=0,\ldots,\deg \}$ and let  points $(i,-\infty)$ lie
below any line in the plane.
Given an index $s$, the following proposition bounds 
the slope of the edge of $CH$ above 
$\deg+1-s$ in terms of the slope 
of the edge of $\app{CH}$ above $\deg+1-s$ and $M$.

\begin{Proposition}
\label{prop_boundslope_int}
\RI{
Given an integer $s$, let 
$t,h,t'$, and $h'$ be integers such that
}
 \begin{enumerate}
  \item[$(i)$] $t<\deg+1-s\leq t+h \leq \deg$, 
  \item[$(ii)$] $\forall 0\leq i \leq \deg$,
 the point $(i,\app{p_i})$
 lies below the line
 $((t,\app{p_{t}}), (t+h,\app{p_{t+h}}))$,
  \item[$(i')$] $t'<\deg+1-s\leq t'+h' \leq \deg$, and
  \item[$(ii')$] $\forall 0\leq i \leq \deg$,
 the points $(i,p_i)$
 lie below the line
 $((t',p_{t'}), (t'+h',p_{t'+h'}))$.
 \end{enumerate}
 Then
  \begin{equation}
 \label{eq_boundslope_int}
  \dfrac{\app{p_{t+h}}-\app{p_{t}}}{h} - 2^{-M+1} \leq
  \dfrac{p_{t'+h'}-p_{t'}}{h'} \leq 
  \dfrac{\app{p_{t+h}}-\app{p_{t}}}{h} + 2^{-M+1}.
 \end{equation}
\end{Proposition}

\RI{For a given integer $s$, 
the existence of integers $t,h,t',h'$
satisfying $(i)$, $(ii)$, $(i')$, $(ii')$
follows from the existence of the convex hulls
$CH$ and $\app{CH}$.}
We postpone the proof of Prop.~\ref{prop_boundslope_int}.
Remark that $2^{2^{-L}} \leq 1+2^{-L}$
for  $L\geq 0$,
apply Prop.~\ref{prop_edgeofCH_int}, and
obtain:

\begin{Corollary}[of Prop.~\ref{prop_boundslope_int}]
 \label{cor_edgeofCHInterval_int}
 Let $s,t,h$ be as in Prop.~\ref{prop_boundslope_int}.
 Define $\app{\rho_s}'$ as
 $\left|\dfrac{P_t}{P_{t+h}}\right|^{\frac{1}{h}}$.
 Then
  \begin{equation}
 \label{eq_boundrr_int}
        \frac{\app{\rho_s}'}{(2\deg)(1+2^{-M+1})} < \rootrad_s(\pol) 
    < (2\deg)(1+2^{-M+1})\app{\rho_s}'.
 \end{equation}
\end{Corollary}

We are ready to describe our Algo.~\ref{algo:taskSprime}, which solves the 
\RRCstar problem.
In steps 1-2, 
 $1$-bit approximations $\app{p_i}$ of $p_i=\log|\pol_i|$
 are computed from $\pol_i$, for $i=0,\ldots,\deg$.
This requires $O(d\log\|\pol\|)$
bit operations.
In step 3 we compute  the convex hull $\app{CH}$ of a polygon with $\deg+1$ vertices
$(0,\app{p_0}),\ldots, (\deg,\app{p_\deg})$ ordered with respect to their ordinates.
Using  Graham's algorithm of  \cite{graham1983finding}, 
we only need $O(\deg)$ arithmetic operations
(additions)
with numbers of magnitude $O(\log\|P\|)$.
In steps 4,5,6, the $\app{\rho_s}'$'s
for $s=0,\ldots,\deg$
are computed as in Cor.~\ref{cor_edgeofCHInterval_int}.
This task is performed with  rounding 
to double precision arithmetic  and requires  $O(\deg)$ bit operations.
Finally, 
$(1+2^{-M+1})\leq 2$ if $M\geq 1$;
thus the $\app{\rho_s}'$'s in the output
satisfy 
$
		 \forall s=1,\ldots,\deg,~~\frac{\app{\rho_s}'}{4\deg} < \rootrad_s(\pol) 
		                     < (4\deg)\app{\rho_s}',
		$
and 
 Prop.~\ref{prop_tasksp}
follows.

\begin{algorithm}[t!]
	\begin{algorithmic}[1]
	\caption{$\solveRRCstar(\pol)$}
	\label{algo:taskSprime}
	\Require{$P\in\G[\var]$ of degree $\deg$ s.t. $P(0)\neq 0$.
    } 
	\Ensure{ 
            $d$ positive real numbers 
            $\app{\rho_1}',\ldots,\app{\rho_\deg}'$.
	}
	\For { $i=0,\ldots,\deg$}
        \State Compute $\app{p}_i$, a 1-bit approximation of $p_i$, as defined in Eq.~(\ref{eq_appi}).
	\EndFor
	\State $\app{CH}\ass \{(i_k,\app{p_{i_k}}) | k=0,\ldots,\ell\}$, the upper part of the convex hull of $\{ (i,\app{p}_i) | i = 0,\ldots, \deg \}$
	\For { $k=1,\ldots,\ell$ }
        \For {$s=\deg+1-i_k,\ldots,\deg+1-i_{k-1}$}
            \State $\app{\rho_{s}}'\ass | \frac{\pol_{i_{k-1}}}{\pol_{i_{k}}} |^{\frac{1}{i_k-i_{k-1}}}$
            {\it //double precision floating point}
        \EndFor
	\EndFor
    \State \Return $\app{\rho_1}',\ldots,\app{\rho_\deg}'$
	\end{algorithmic}
\end{algorithm}
	
\subsection{Proof of Prop.~\ref{prop_boundslope_int}}
\label{subsec_proof_prop}

\begin{figure}
	 \centering
	  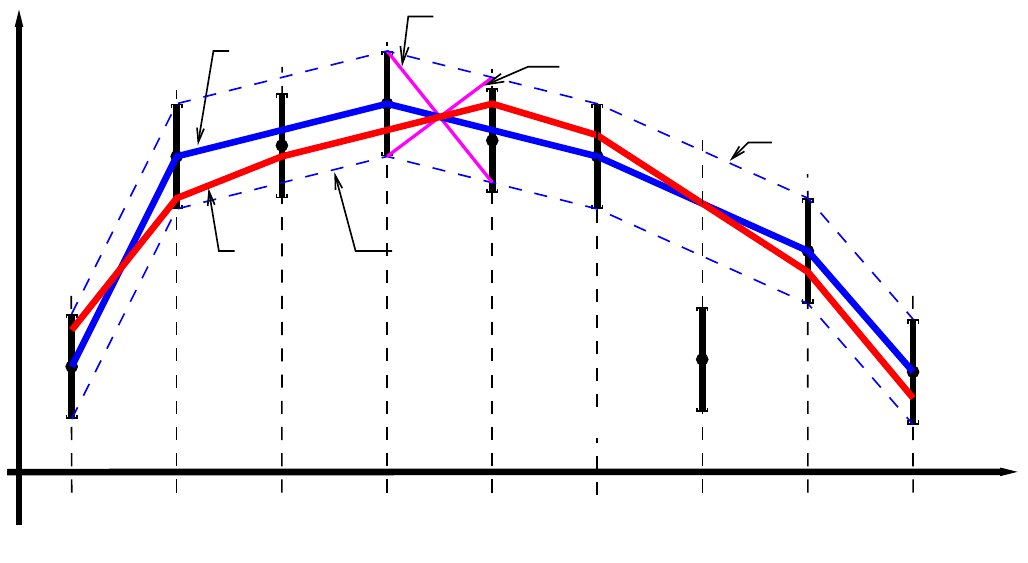
	    \caption{The convex hulls $CH$ and $\app{CH}$.}
	 \label{fig:convexHull}
	 \vspace*{-0.5cm}
	\end{figure}

For $i=0,\ldots, \deg$, define
$\app{p_i}^+$ and $\app{p_i}^-$
as
\[
 \app{p_i}^+ = \left\{
 \begin{array}{ll}
  \app{p_i} + 2^{-M} & \text{ if } |\app{p_i}|> -\infty,\\
  -\infty          & \text{otherwise}
 \end{array}
 \right.
 \text{, and }
 \app{p_i}^- = \left\{
 \begin{array}{ll}
  \app{p_i} - 2^{-M} & \text{ if } |\app{p_i}|> -\infty,\\
  -\infty          & \text{otherwise}
 \end{array}
 \right.
 .
\]
$\app{CH}$ is the upper part of the convex hull
of $\{ (i,\app{p_i}) | i=0,\ldots,\deg \}$.
Suppose that it is the poly-line passing through 
$\{ (i_k,\app{p_{i_k}}) | k=0,\ldots,\ell \}$.
It defines two poly-lines:
\begin{itemize}
 \item $\app{CH}^+$, the poly-line with vertices 
       $\{ (i_k,\app{p_{i_k}}^+ | k=0,\ldots,\ell \}$, and
 \item $\app{CH}^-$, the poly-line  with vertices 
       $\{ (i_k,\app{p_{i_k}}^-) | k=0,\ldots,\ell \}$.
\end{itemize} 
$CH$ is the upper part of the convex hull of $\{ (i,p_i) | i=0,\ldots,\deg \}$, and suppose it
is the poly-line with vertices 
$\{ (j_k,p_{j_k}) | k=0,\ldots,\ell' \}$.
For demonstration see Fig.~\ref{fig:convexHull}
where $\deg=8$, 
the $\app{p_i}$'s are drawn with black
circles, the intervals $[\app{p_i}^-,\app{p_i}^+]$ with bold vertical bars,
$\app{CH}$ with a bold blue poly-line,
$\app{CH}^+$ and $\app{CH}^-$ with dashed blue poly-lines,
and $CH$ with a bold red line.
One has:
\begin{Proposition}
\label{prop_CH_bound_int}
 The poly-line $CH$ lies below the poly-line $\app{CH}^+$
 and above the poly-line $\app{CH}^-$.
\end{Proposition}

\noindent
{\bf Proof of Prop.~\ref{prop_CH_bound_int}:}
In order to prove that $CH$ lies below $\app{CH}^+$,
we show that 
if $j_t, i_k, i_{k'}$ is a triple
of integers 
such that $(j_t,p_{j_t})$ is a vertex of $CH$ and 
$[ (i_k,\app{p_{i_k}}^+), (i_{k'},\app{p_{i_{k'}}}^+) ]$
is an edge of $\app{CH}^+$,
then
$(j_t,p_{j_t})$ 
lies on or below
the line $( (i_k,\app{p_{i_k}}^+), (i_{k'},\app{p_{i_{k'}}}^+) )$.
Suppose 
this is not the case, \emph{i.e.} 
the point
$(j_t,p_{j_t})$ lies strictly above 
the line
$( (i_k,\app{p_{i_k}}^+), (i_{k'},\app{p_{i_{k'}}}^+) )$.
Since $p_{j_t}\leq \app{p_{j_t}}^+$,
$\app{p_{j_t}}^+$ lies strictly above 
$( (i_k,\app{p_{i_k}}^+), (i_{k'},\app{p_{i_{k'}}}^+) )$, 
thus
$\app{p_{j_t}}$ lies strictly above 
$( (i_k,\app{p_{i_k}}), (i_{k'},\app{p_{i_{k'}}}) )$ 
and
 $\app{CH}$ is not the convex hull of 
$\{ (i,\app{p_i}) | i=0,\ldots,\deg \}$, which is a contradiction.

In order to show that $\app{CH}^-$ lies below $CH$,
we show that for a given triple of integers $i_t, j_k, j_{k'}$
such that $(i_t,\app{p_{i_t}}^-)$ is a vertex of $\app{CH}^-$ and 
$[ (j_k,p_{j_k}), (j_{k'},p_{j_{k'}}) ]$
is an edge of $CH$, 
the point
$(i_t,\app{p_{i_t}}^-)$
lies on or below
the line $( (j_k,p_{j_k}), (j_{k'},p_{j_{k'}}) )$.
Suppose it is not the case.
Since $p_{i_t}\geq \app{p_{i_t}}^-$,
the point
$p_{i_t}$ lies strictly above
the line passing through $( (j_k,p_{j_k}), (j_{k'},p_{j_{k'}}) )$
and $CH$ is not the convex hull of 
$\{ (i,p_i) | i=0,\ldots,\deg \}$, which is a contradiction.
\qed
\medskip

\noindent
{\bf Proof of Prop.~\ref{prop_boundslope_int}:}
Given the integer $s$, 
 let $t,h,t',h'$ be integers such that
 conditions $(i), (ii), (i')$ and $(ii')$ hold.

By virtue of  $(i')$ and $(ii')$, 
 $
](t',p_{t'}), (t'+h',p_{t'+h'})]$
 is the edge of $CH$ whose orthogonal projection
 onto the abscissa axis contains $\deg+1-s$,
 and $\dfrac{p_{t'+h'} - p_{t'}}{h'}$ is the slope
 of that edge.
 
By virtue of  $(i)$ and $(ii)$,
 $](t,\app{p_{t}}), (t+h,\app{p_{t+h}})]$
 is the edge of $\app{CH}$
 whose orthogonal projection
 onto the abscissa axis contains $\deg+1-s$.
 Consider the two segments
 $](t,\app{p_{t}}^-), (t+h,\app{p_{t+h}}^-)]$
 and 
 $](t,\app{p_{t}}^+), (t+h,\app{p_{t+h}}^+)]$
 that are the edges of $\app{CH}^-$ and $\app{CH}^+$,  respectively,
 whose orthogonal projections
 onto the abscissa axis also contain $\deg+1-s$.

 From Prop.~\ref{prop_CH_bound_int} $CH$ is a poly-line enclosed by $\app{CH}^-$ and $\app{CH}^+$ and since the first coordinates of its
 vertices are integers, 
 its slope $\dfrac{p_{t'+h'} - p_{t'}}{h'}$
 above $\deg+1-s$ is bounded below by 
 $\dfrac{\app{p_{t+h}} - \app{p_{t}}}{h} - 2^{-M+1}$
 and above by
 $\dfrac{\app{p_{t+h}} - \app{p_{t}}}{h} + 2^{-M+1}$,
which proves Prop.~\ref{prop_boundslope_int}.
See Fig.~\ref{fig:convexHull} for an illustration.
\qed

\subsection{Solving the \RRC problem}
\label{subsec_tasks}

\begin{algorithm}[t!]
	\begin{algorithmic}[1]
	\caption{$\solveRRC(\pol,c,\delta)$}
	\label{algo:taskS}
	\Require{
            $\pol\in\Z[\var]$ of degree $\deg$,
            a center $c\in\G$ and a relative precision $\delta>0$.
    } 
	\Ensure{ 
            $d$ positive real numbers 
            $\rho_{c,s},\ldots,\rho_{c,\deg}$ solving task S.
	}
	\State $g\ass \lceil \log \dfrac{\log(4\deg)}{\log(1+\delta)}\rceil$
	\State compute $\Graeffe{\shift{\pol}{c}}{g}$
	\State $\rho_1',\ldots,\rho_{\deg}'\ass \solveRRCstar(\Graeffe{\shift{\pol}{c}}{g})$
	\For { $s=0,\ldots,\deg$}
        \State $\rho_{c,s}\ass (\rho_s')^{1/2^{g}}$ 
	\EndFor
    \State \Return $\rho_{c,1},\ldots,\rho_{c,\deg}$
	\end{algorithmic}
\end{algorithm}

Using Rem.~\ref{rem_numofGraeffeItts_app}, we define in Algo.~\ref{algo:taskS} the algorithm $\solveRRC$.
To estimate the cost at steps 2 and 3,
let $\mathcal{M}:\N\rightarrow \N$ be
such that
two polynomials of degree at most $\deg$
and bit-size at most $\tau$
can be multiplied
by using $O(\mathcal{M}(\deg\tau))$
bit operations.
Recall the following:
\begin{enumerate}[$1.$]
 \item computing $\shift{\pol}{c}$ requires 
       $O(\mathcal{M}(\deg^2\log\deg + \deg^2\log(|c|+1) + \deg\log\|\pol\|))$
       bit operations, 
 \item $\|\shift{\pol}{c}\|\leq \|\pol\|(|c|+1)^\deg$,
 \item computing $\|\Graeffe{\shift{\pol}{c}}{i}\|$ from 
                 $\|\Graeffe{\shift{\pol}{c}}{i-1}\|$
       requires $O(\mathcal{M}(\deg\log\|\Graeffe{\shift{\pol}{c}}{i-1}\|))$
       bit operations,
 \item $\|\Graeffe{\shift{\pol}{c}}{i}\| \leq (\deg+1)(\|\Graeffe{\shift{\pol}{c}}{i-1}\|)^2\leq \ldots \leq (\deg+1)^{2^i}(\|\shift{\pol}{c}\|)^{2^i}$.
\end{enumerate}
For 1 and 2, see for instance \cite{von1997fast}[Theorem 2.4]
and \cite{von1997fast}[Lemma 2.1]).
3 and 4 are derived
from Eqs.~(\ref{eq_compDLG}) and~(\ref{eq_defbk}), respectively.
From 2, 4 and  $g\in O(\log\deg)$ one obtains 
\begin{equation}
\label{eq_bs}
 \log\| \Graeffe{\shift{\pol}{c}}{g} \|
 \in O(\deg\log(\deg+1) + \deg\log\|\pol\| + \deg^2\log(|c|+1)),
\end{equation}
thus performing $g$ DLG iterations for $\shift{\pol}{c}$
involves
\[ 
 O(g\mathcal{M}(\deg\log\| \Graeffe{\shift{\pol}{c}}{g} \|))
 = O(\log\deg \mathcal{M}(\deg(\deg\log(\deg+1) + \deg\log\|\pol\| + \deg^2\log(|c|+1))))
\] bit operations;
 this
dominates the cost of step 2.
Due to Sch\"{o}nhage-Strassen or Harvey-van der Hoeven multiplication, 
$\mathcal{M}(n)\in\app{O}(n)$, and so step 2 involves
\[
 \app{O}( \deg^2\log\|\pol\| + \deg^3\log(|c|+1))
\] bit operations.
Step 3 involves
 $O(\deg \log\| \Graeffe{\shift{\pol}{c}}{g} \|)$ bit operations, 
the cost of the {\bf for} loop in steps 4-5
is dominated by
the cost of step 2, and we
 complete the proof of Thm.~\ref{Th_solving_taskS}.

\subsection{Implementation Details}
\label{subsection_impl}

The exact computation of
$\Graeffe{\shift{\pol}{c}}{g}$  
can involve numbers of
very large bit-size (see eq.~(\ref{eq_bs})), and the 
key point for the practical efficiency of our implementation of 
Algo.~\ref{algo:taskS} is to avoid this.
Instead, we use \emph{ball arithmetic}, \emph{i.e.}
arbitrary precision floating point arithmetic with absolute error bounds,
implemented for instance in the \clang library {\tt arb} 
(see \cite{Johansson2017arb}).

\section{Real Root Isolation}
\label{section_RRI}

In this section, we approximate the root distances to $0$ in order to
improve in practice subdivision approaches to 
real root isolation, and in particular the one described in 
Subsec.~\ref{subsec_subdivision}.
Let us define the notion of annuli cover of the roots
of $\pol\in\Z[\var]$.

\begin{definition}
 A set $\annset_c$ of disjoint concentric annuli centered in $c$ is an \emph{annuli cover} of the roots 
 of $\pol$ of degree $\deg$ if
 \begin{enumerate}
  \item $\forall A\in \annset_c$,
  there are integers $t(A)$ and $h(A)$
  such that 
  \[
 \alpha_{t(A)}(\pol,c),\alpha_{t(A)+1}(\pol,c),\ldots,\alpha_{t(A)+h(A)}(\pol,c)\in A,
\]
 \item $\forall i\in \{1,\ldots,\deg\}$,
 there is an $A\in \annset_c$ such that $\alpha_i(\pol,c)\in A$.
 \end{enumerate}

 For an annulus $A\in\annset_c$,
 $\underline{r}(A)$ and $\overline{r}(A)$
 are the interior and exterior radii of $A$, respectively.
 Write $s_+(A):=\pol(c+\underline{r}(A))\pol(c+\overline{r}(A))$
 and 
 $s_-(A):=\pol(c-\underline{r}(A))\pol(c-\overline{r}(A))$.
\end{definition}
Given an annuli cover $\annset_0$ centered at $0$, we can skip many calls for
 exclusion and root-counting
based on the following:
\begin{Remark}
 \label{remark_counting}
 Let $\ann\in\annset_0$ such that $\underline{r}(A)>0$.
 \begin{enumerate}
  \item If $h(A)=0$ and $s_+(A)>0$ (resp. $s_-(A)>0$),
        $A\cap\R_+$ (resp. $A\cap\R_-$) contains no real root of $\pol$.
  \item If $h(A)=0$ and $s_+(A)<0$ (resp. $s_-(A)<0$),
        $A\cap\R_+$ (resp. $A\cap\R_-$) contains one real root of $\pol$.
  \item If $h(A)>0$ and $s_+(A)<0$ (resp. $s_-(A)<0$),
        $A\cap\R_+$ (resp. $A\cap\R_-$) contains at least one real root of $\pol$.
 \end{enumerate}
\end{Remark}

In Subsections \ref{subsec_exclTest} and~\ref{subsec_countTest} 
we describe our exclusion test and root counter
based on Rem.~\ref{remark_counting}.
In Subsec.~\ref{subsec_algoRRI}
we describe our algorithm solving the RRI 
problem. In Subsec.~\ref{subsec_expeRes_real}
we 
present the results
of our  numerical tests.

\subsection{Annuli cover and exclusion test}
\label{subsec_exclTest}

 \begin{algorithm}[t!]
	\begin{algorithmic}[1]
	\caption{$\CRzero{B, \pol,\annset}$}
	\label{algo:excl_RRI}
	\Require{
            A polynomial $\pol\in\Z[\var]$ of degree $\deg$,
            a segment $B$ of $\R$,
            an annuli cover $\annset$ of the roots of $\pol$ centered in $0$.
            Assume $0\notin B$.
    } 
	\Ensure{ 
            an integer in $\{ -1,0\}$;
            if $0$ then $P$ has no real root in $B$;
            if $-1$ then there is a root in $2B$.
	}
	\State Compute $n(\annset,B)$, $n_0(\annset,B)$ and 
	       $n_{\geq1}(\annset,B)$
	\If { $n(\annset,B)=n_0(\annset,B)$ } 
        \State \Return $0$
    \EndIf
    \If { $n_{\geq1}(\annset,B)>=1$ }
        \State \Return $-1$
    \EndIf
    \State \Return $\Tzero{\contDisc{B}, \pol}$
	\end{algorithmic}
	\end{algorithm}
	
 Let $B$ be a real line segment that does not contain $0$, let
 $\annset$ be an annuli cover centered in $0$, and let
 $A\in\annset$. Define that:
\begin{itemize}
 \item $s_B$, the sign of $B$, is $-1$ (resp. $1$) if $B<0$ (resp. $B>0$),
 \item $\R s_B(A)$ is $A\cap\R_-$ if $s_B<0$, and is $A\cap\R_+$ otherwise,
 \item $ss_B(A)$ is $s_-(A)$ if $s_B<0$, and is $s_+(A)$ otherwise,
  \item $n(\annset,B)$ is the number of annuli $A\in\annset$ s.t. $A\cap B\neq \emptyset$,
  \item $n_0(\annset,B)$ is the number of annuli $A\in\annset$ s.t. 
  \[
   (A\cap B\neq \emptyset)\wedge(h(A)=0)\wedge ss_B(A)>0,
  \]
  \item $n_{\geq1}(\annset,B)$: the number of annuli $A\in\annset$ s.t. 
  \[
   (A\cap B\neq \emptyset)\wedge(h(A)\geq0)\wedge ss_B(A)<0 
   \wedge \R s_B(A)\subseteq 2B.
  \]
 \end{itemize}
By virtue of  Rem.~\ref{remark_counting},
 if $n(\annset,B)=n_0(\annset,B)$,  
 then all the annuli intersecting $B$ contain no root, thus $B$ contains no root.
 If $n_{\geq1}(\annset,B)\geq 1$ then $2B$ contains at least one real root.
 
 Our exclusion test $\CRzerot$
 is described in Algo.~\ref{algo:excl_RRI}.
 For computing $n(\annset,B)$, $n_0(\annset,B)$ and 
$n_{\geq1}(\annset,B)$ in Step 1, we use double precision interval arithmetic,
hence
Step 1 involves $O(\deg)$  bit operations.
This implies 
\begin{Proposition}
  \label{prop_CzeroCostR}
 Let $B$ be a real line segment with $0\notin B$,
 and let $\annset$ be an annuli cover of the roots of $\pol$ centered in $0$.
 The cost of carrying out $\CRzero{B, \pol,\annset}$ is bounded by the cost of carrying out $\Tzero{\contDisc{B}, \pol}$.
 
 $\CRzero{B,\pol,\annset}$ returns an integer 
 $k$ in $\{-1,0\}$.
 If $k=0$, then $\pol$ has no real roots in $B$.
 If $k=-1$, then $\pol$ has a root in $2B$.
\end{Proposition}

\subsection{Annuli cover and root counter}
\label{subsec_countTest}

In order to describe our root counter,
we define:
\begin{itemize}
  \item $n_1(\annset,B)$: the number of annuli $A\in\annset$ s.t. :
  \[
   (A\cap B\neq \emptyset)\wedge(h(A)=0)\wedge (ss_B(A)<0) \wedge 
   (\R s_B(A)\subseteq B),
  \]
  \item $n_{\geq1}'(\annset,B)$: the number of annuli $A\in\annset$ s.t.:
  \[
   (~A\cap B\neq \emptyset)\wedge(h(A)\geq0)\wedge ss_B(A)<0 
   \wedge (\R s_B(A)\subseteq 2B\setminus(1/2)B~).
  \]
 \end{itemize}
By virtue of Rem.~\ref{remark_counting},
 if $n(\annset,B)=n_0(\annset,B)+n_1(\annset,B)$,
$B$ contains exactly $n_1(\annset,B)$ real roots.
If $n_{\geq1}'(\annset,B)\geq 1$ then $\pol$
has at least one real root in $2B\setminus(1/2)B$.

\begin{algorithm}[t!]
	\begin{algorithmic}[1]
	\caption{$\CRstar{B,\pol,\annset}$}
	\label{algo:count_RRI}
	\Require{
            A polynomial $\pol\in\Z[\var]$ of degree $\deg$,
            a segment $B$ of $\R$,
            an annuli cover $\annset$ of the roots of $\pol$ centered in $0$.
            Assume $0\notin 2B$.
    } 
	\Ensure{ 
            an integer $k$ in $\{ -1,0,1,\ldots,d\}$;
            if $k\geq0$ then $P$ has $k$ roots in $B$;
            if $-1$ then there is a root in $2B\setminus(1/2)B$.
	}
	\State Compute $n(\annset,B)$, $n_0(\annset,B)$, $n_1(\annset,B)$ and 
	       $n_{\geq1}'(\annset,B)$
	\If { $n(\annset,B)=n_0(\annset,B)+n_1(\annset,B)$ } 
        \State \Return $n_1(\annset,B)$
    \EndIf
    \If { $n_{\geq1}'(\annset,B)\geq 1$ }
        \State \Return $-1$
    \EndIf
    \State \Return $\Tstar{\contDisc{B}, \pol}$
	\end{algorithmic}
	\end{algorithm}

Our root counter is described in 
Algo.~\ref{algo:count_RRI}.
For computing $n(\annset,B)$, $n_0(\annset,B)$, $n_1(\annset,B)$ and 
$n_{\geq1}'(\annset,B)$ in Step 1, we use double precision interval arithmetic,
thus step 1 involves $O(\deg)$  bit operations. 

\begin{Proposition}
  \label{prop_CstarCostR}
 Let $B$ be a real line segment with $0\notin B$ 
 and let $\annset$ be an annuli cover of the roots of $\pol$ centered in $0$.
 The cost of carrying out $\CRstar{B, \pol,\annset}$ is bounded by the cost of carrying out $\Tstar{\contDisc{B}, \pol}$.
 
 $\CRstar{B,\pol,\annset}$ returns an integer 
 $k$ in $\{-1,0,\ldots,\deg\}$.
 If $k\geq 0$, then $\pol$ has $k$ roots in $\Delta(B)$.
 If $k=-1$, then $\pol$ has a root in $2B\setminus (1/2)B$.
 \end{Proposition}

\subsection{Annuli cover and the RRI problem}
\label{subsec_algoRRI}

Consider the following procedure.
\medskip

\noindent{\bf Stage 1:} Compute $\annset_0$ 
                        by calling $\solveRRC(\pol,0,d^{-2})$.\\
\noindent{\bf Stage 2:} Apply the subdivision procedure of Subsec.~\ref{subsec_subdivision}
                 while using $\CRzero{B, \pol,\annset_0}$ 
                 (resp. $\CRstar{B, \pol,\annset_0}$)
                 as an exclusion test
                 (resp. root counter)
                 for real line segment
                 $B$ of the subdivision tree.
                 In the verification step of Newton iterations, use the 
                 $\Tstart$-test of \cite{becker2018near}.
\medskip

At Stage 1, we obtain $\annset_0$ by computing the 
connected components made up of the concentric annuli defined by the output
of $\solveRRC(\pol,0,d^{-2})$.

By virtue  of Thm.~\ref{Th_solving_taskS} and
Propositions ~\ref{prop_CzeroCostR} and ~\ref{prop_CstarCostR}, 
this procedure solves the RRI problem, and 
its bit-complexity 
is bounded by the bit-complexity 
of the algorithm described in \cite{becker2016complexity},
thus it is near optimal for the benchmark problem.

\subsection{Experimental results}
\label{subsec_expeRes_real}

The procedure given in Subsec.~\ref{subsec_algoRRI}
has been implemented within the library \ccluster;  we
call this implementation \risolateRR. 
Comparison of \risolateRR with \risolate reveals
 practical improvement due to 
using our root radii 
algorithms in
 subdivision process. 
We also compare \risolateRR
with the subdivision algorithm
of \cite{sagraloff2016computing}
whose implementation \anewdsc is described in 
\cite{Kobel} and is currently the user's choice for 
real root isolation.

\subsubsection{Test polynomials}
\label{subsubsec_testSuite}
We consider the following polynomials.

The Bernoulli polynomial of degree $\deg$ is
$B_\deg(z)=\sum_{k=0}^{\deg} {{\deg}\choose{k}}b_{\deg-k}z^k$
where the $b_i$'s are the Bernoulli numbers.

The Wilkinson polynomial of degree $\deg$
is $W_\deg(z) = \prod_{i=1}^\deg(z-i)$.

For integers $n>0$, we define polynomials
with $(2n+1)\times(2n+1)$ roots on 
the nodes of a regular grid centered at $0$
as
\[P_{(2n+1)\times(2n+1)}(z)=\prod_{-n\leq a,b \leq n}(z-a+\ii b).\]

The Mignotte polynomial of degree $d$
and bitsize $\tau$ is
$M_{d,\tau}(z)=z^d - 2(2^{\frac{\tau}{2}-1}z-1)^2$.

We also consider dense polynomials
 of degree $d$
with coefficients randomly chosen within $[-2^{\tau-1}, 2^{\tau-1}]$ (under the uniform distribution).

\subsubsection{Results}
\label{subsubsec_results_real}

In our test polynomials 
with several degrees/bit-sizes, 
we used \risolate, \risolateRR and \anewdsc
to solve the RRI problem.
Our non-random examples have only simple roots
and for those examples
\anewdsc is called with option {\tt -S 1} to avoid  testing input polynomial for being square-free.

Times are sequential times
in seconds on a \machine machine with Linux.
We report in Tab.~\ref{table_RRIAlgo}:\\
\noindent - $d$, $\tau$ and $d_\R$, that is, the degree, the bit-size and the number of real roots, respectively,
\\
\noindent - $t_1$ (resp. $t_2$), the running time of \risolate (resp. \risolateRR),\\
\noindent - $n_1$ (resp. $n_2$), the number of $\Tzerot$-tests in \risolate (resp. \risolateRR),\\
\noindent - $n_1'$ (resp. $n_2'$), the number of $\Tstart$-tests in \risolate (resp. \risolateRR),\\
\noindent - $t_3$, the time required
       to compute the annuli cover in \risolateRR,\\
\noindent - $t_4$, the running time in second of \anewdsc.\\
For random polynomials, we display averages over 10 examples 
of those values. We also display $\sigma_1$, $\sigma_2$,
and $\sigma_4$, the standard deviation of running time
of  \risolate, \risolateRR and \anewdsc, respectively.

\begin{table}[t!]
\centering
\begin{scriptsize}
 \begin{tabular}{@{}r@{\hspace{2mm}}r@{\hspace{2mm}}r||c@{\hspace{2mm}}c||c@{\hspace{2mm}}c@{\hspace{1mm}}c@{\hspace{1mm}}||c||c@{\hspace{1mm}}}
       &     &          & \multicolumn{2}{c||}{\risolate}  
                        & \multicolumn{4}{c||}{\risolateRR}
                        & \multicolumn{1}{c}{\anewdsc}\\\hline
$d$ & $\tau$ & $d_\R$   & $t_1$ $(\sigma_1)$ & $n_1,n_1'$
                        & $t_2$ $(\sigma_2)$ & $n_2,n_2'$ & $t_3/t_2$ (\%)
                        & $t_2/t_1$ (\%)
                        & $t_4$ $(\sigma_4)$ \\\hline
\multicolumn{10}{c}{10 monic random dense polynomials per degree/bit-size} \\\hline
256 & 8192  & 6.00 & 3.02 (1.13) & 128.,80.2 & \coblue{.374} (.080) & 4.40,25.3 & 16.3 & 12.3 & \cored{.784} (1.73)\\
256 & 16384 & 7.80 & 5.09 (1.99) & 183.,122. & \coblue{.499} (.132) & 2.60,22.9 & 22.2 & 9.80 & \cored{2.76} (5.19)\\
256 & 32768 & 7.40 & 7.59 (3.20) & 172.,125. & \coblue{.442} (.174) & 4.40,27.4 & 16.3 & 5.82 & \cored{1.18} (.600)\\
256 & 65536 & 7.00 & 10.7 (6.33) & 170.,140. & \coblue{.480} (.160) & 4.30,25.4 & 10.4 & 4.46 & \cored{1.91} (1.18)\\\hline
391 & 8192  & 7.20 & 8.87 (2.99) & 157.,107. & \coblue{1.12} (.310) & 4.60,26.0 & 15.0 & 12.6 & \cored{3.29} (5.42)\\
391 & 16384 & 8.40 & 10.1 (4.12) & 186.,116. & \coblue{1.39} (.575) & 6.20,28.9 & 15.3 & 13.7 & \cored{10.2} (19.8)\\
391 & 32768 & 8.60 & 18.6 (6.98) & 202.,155. & \coblue{1.38} (.528) & 4.00,29.0 & 14.5 & 7.41 & \cored{1.67} (.750)\\
391 & 65536 & 7.60 & 23.9 (13.9) & 178.,137. & \coblue{1.88} (1.17) & 3.90,33.6 & 18.5 & 7.86 & \cored{13.9} (18.9)\\\hline
512 & 8192  & 6.60 & 31.1 (18.5) & 158.,104. & \cored{3.68} (4.72) & 6.00,25.9 & 12.4 & 11.8 & \coblue{1.26} (1.03)\\
512 & 16384 & 5.20 & 41.1 (20.1) & 152.,106. & \cored{5.00} (4.63) & 6.50,25.8 & 5.37 & 12.1 & \coblue{1.70} (2.17)\\
512 & 32768 & 6.00 & 56.7 (28.1) & 167.,122. & \coblue{2.00} (.596) & 4.40,28.4 & 18.1 & 3.53 & \cored{5.95} (7.61)\\
512 & 65536 & 6.60 & 86.5 (34.2) & 180.,137. & \coblue{4.84} (3.67) & 5.90,32.7 & 5.19 & 5.60 & \cored{60.1} (118.)\\\hline

\multicolumn{10}{c}{Bernoulli polynomials} \\\hline
256  & 1056 & 64  & 1.13 & 292, 82   & \coblue{0.08} & 12, 3 & 54.2 & 7.77 & \cored{0.20}\\
391  & 1809 & 95  & 2.66 & 460, 145  & \coblue{0.30} & 12, 2 & 76.1 & 11.2 & \cored{1.09}\\
512  & 2590 & 124 & 6.15 & 528, 144  & \coblue{0.38} & 14, 3 & 65.9 & 6.30 & \cored{1.58}\\
791  & 4434 & 187 & 16.3 & 892, 264  & \coblue{2.39} & 20, 1 & 85.0 & 14.6 & \cored{9.92}\\
1024 & 6138 & 244 & 56.3 & 1048, 283 & \coblue{2.42} & 12, 3 & 76.5 & 4.30 & \cored{14.9}\\\hline
\multicolumn{10}{c}{Wilkinson polynomials} \\\hline
256  & 1690 & 256  & 3.63 & 1030, 283  & \coblue{0.17} & 0, 10 & 41.1 & 4.90 & \cored{1.57}\\
391  & 2815 & 391  & 17.6 & 1802, 541  & \coblue{0.68} & 0, 10 & 51.7 & 3.88 & \cored{5.69}\\
512  & 3882 & 512  & 25.9 & 2058, 533  & \coblue{1.04} & 0, 11 & 46.9 & 4.01 & \cored{27.1}\\
791  & 6488 & 791  & 165. & 3698, 1110 & \coblue{7.04} & 0, 11 & 57.1 & 4.26 & \cored{158.}\\
1024 & 8777 & 1024 & 265. & 4114, 1049 & \coblue{8.38} & 0, 12 & 51.2 & 3.15 & \cored{309.}\\\hline
\multicolumn{10}{c}{Polynomials with roots on a regular grid} \\\hline
289  & 741  & 17 & 0.40 & 86, 30  & \cored{0.13} & 0, 16 & 81.7 & 34.4 & \coblue{0.09}\\
441  & 1264 & 21 & 0.91 & 106, 36 & \coblue{0.21} & 0, 20 & 77.3 & 23.4 & \cored{0.39}\\
625  & 1948 & 25 & 1.59 & 118, 39 & \cored{0.92} & 0, 24 & 89.1 & 58.0 & \coblue{0.80}\\
841  & 2800 & 29 & 3.30 & 154, 51 & \coblue{1.67} & 0, 28 & 87.4 & 50.7 & \cored{2.56}\\
1089 & 3828 & 33 & 8.06 & 166, 55 & \coblue{2.20} & 0, 32 & 76.4 & 27.3 & \cored{4.49}\\\hline
\multicolumn{10}{c}{Mignotte polynomials} \\\hline
512 & 256  & 4 & 1.57 & 34, 15 & \cored{1.67} & 2, 12 & 16.5 & 106. & \coblue{0.76}\\
512 & 512  & 4 & 3.07 & 34, 15 & \cored{4.81} & 2, 14 & 5.70 & 156. & \coblue{1.90}\\
512 & 1024 & 4 & 5.91 & 34, 15 & \cored{5.96} & 2, 10 & 4.13 & 100. & \coblue{5.28}\\
512 & 2048 & 4 & 13.8 & 34, 15 & \coblue{13.2} & 2, 9  & 2.42 & 95.3 & \cored{14.1}\\
512 & 4096 & 4 & 29.7 & 50, 17 & \coblue{30.8} & 2, 6  & .753 & 103. & \cored{36.0}\\
\hline

 \end{tabular}
 \caption{Runs of \risolate, \risolateRR and \anewdsc on our test polynomials.}
\label{table_RRIAlgo}
\end{scriptsize}
\vspace*{-1cm}
\end{table}

Compare columns $n_1,n_1'$ and $n_2,n_2'$ in Tab.~\ref{table_RRIAlgo}:
using the annuli cover
both in exclusion tests and root counter
 reduces dramatically the number of Pellet's
tests performed in the subdivision process,
and significantly decreases the running time
(see column $t_2/t_1$).
In the cases 
where the ratio $\tau/\deg$ is low
 \risolateRR  spent most of the time on 
solving the \RRC problem (see column $t_3/t_2$).
Finally, \anewdsc remains faster than 
\risolateRR 
for polynomials having a few real roots or a low
bit-size,
whereas this trend seems to reverse when
the ratios of the number of real roots 
and/or bit-size 
over the degree increase
(see columns $t_2$ and $t_4$).
Mignotte polynomials of even degree have four 
real roots among which two are separated by a distance
way less than the relative size of $\deg^{-2}$, the relative
size of annuli in the computed annuli cover.
In such cases, the knowledge of root radii enables no
 significant improvement
 \RI{because subdivision solvers spend most of their
 running time 
 on performing Newton's iterations that converge to the cluster of two close
 roots, and then on separating the two roots}.

\section{Complex Root Clustering}
\label{section_CRC}

In this section, by approximating the root distances from three centers, namely 
$0$, $1$ and $\ii$ we
improve practical performance of subdivision algorithms for 
complex root clustering.

Using three annuli covers $\annset_0, \annset_1$ and 
$\annset_{\ii}$ of the roots of $\pol$,
one can compute a set $\mathcal{D}$ of $O(\deg^2)$
complex discs containing all the roots of $\pol$,
and then skip expensive Pellet-based exclusion tests of the 
 boxes
 that do not intersect the union of these discs.

In Subsec.~\ref{subsec_exclTest_comp} we describe
an exclusion test using the set $\mathcal{D}$ of discs
containing the roots of $\pol$,
and in Subsec.~\ref{subsec_algoCRC} we present
a procedure solving the CRC with near optimal 
bit complexity.
In Subsec.~\ref{subsec_expeRes_complex} we show experimental results.

\subsection{Annuli cover and exclusion test}
\label{subsec_exclTest_comp}

Let $\mathcal{D}$ be a set of $O(\deg^2)$
complex discs covering all the roots of $\pol$,
\emph{i.e.} any root of $\pol$ is in at least one 
disc in $\mathcal{D}$.
A box $B$ such that $B\cap\mathcal{D}=\emptyset$
cannot contain a root of $\pol$.

We define an exclusion test based on the above consideration,
called $\CCzerot$-test and described in Algo.~\ref{algo:excl_CRC}.
For a box $B$ having a nonempty intersection with the real line,
the number 
$n_{\geq1}(\annset_0,B)$ of annuli intersecting $B$ and containing at least
one real root in $B\cap\R$
is used to save some $\Tzerot$-tests.

\begin{algorithm}[t!]
	\begin{algorithmic}[1]
	\caption{$\CCzero{B, \pol,\mathcal{D}, \annset_0}$}
	\label{algo:excl_CRC}
	\Require{
            A polynomial $\pol\in\Z[\var]$ of degree $\deg$,
            a box $B$ of $\C$,
            a set $\mathcal{D}$ of $O(d^2)$ complex discs covering
            all the roots of $\pol$,
            an annuli cover $\annset_0$ centered in $0$
    } 
	\Ensure{ 
            an integer in $\{ -1,0\}$;
            if $0$ then $P$ has no real root in $B$;
            if $-1$ then there is a root in $2B$.
	}
	\State Compute the number $n$ of discs in $\mathcal{D}$
	       having nonempty intersection with $B$.
    \If { $n=0$ }
        \State \Return 0
    \EndIf
    \If { $B\cap \R\neq\emptyset$ }
        \State Compute $n_{\geq1}(\annset_0,B)$
        \If { $n_{\geq1}(\annset_0,B)>=1$ }
            \State \Return $-1$
        \EndIf
    \EndIf
    \State \Return $\Tzero{\contDisc{B}, \pol}$
	\end{algorithmic}
	\end{algorithm}
	
\begin{Proposition}
 \label{prop_CzeroCost}
 Let $\mathcal{D}$ contain $O(\deg^2)$  discs covering the roots of $\pol$
 and let $\annset_0$ be an annuli cover of the roots of $\pol$ centered in $0$.
 The cost of performing  $\CCzero{B, \pol,\mathcal{D}, \annset_0}$ is bounded by the cost of performing $\Tzero{\contDisc{B}, \pol}$.
 
 $\CCzero{B, \pol,\mathcal{D}, \annset_0}$ returns an integer 
 $k$ in $\{-1,0\}$.
 If $k=0$, then $\pol$ has no root in $B$.
 If $k=-1$, then $\pol$ has a root in $2B$.
\end{Proposition}

\subsection{Annuli cover and the CRC problem}
\label{subsec_algoCRC}

Consider the following procedure.
\medskip

\noindent{\bf Stage 1:}
For $c=0,1,\ii$, 
                 compute $\annset_c$ 
                 by calling $\solveRRC(\pol,c,d^{-2})$.\\
\noindent{\bf Stage 2:} Use $\annset_0$, $\annset_1$ and $\annset_\ii$ to
                 compute a set $\mathcal{D}$ of at most $2\deg^2$ discs
                 covering all roots of $\pol$.\\
\noindent{\bf Stage 3:} Apply the Complex Root Clustering Algorithm
                 of Subsec.~\ref{subsec_subdivision}
                 but let it apply $\CCzero{B, \pol,\mathcal{D},\annset_0}$ instead of $\Tzero{\contDisc{B}, \pol}$
                 as the exclusion test for boxes $B$ of the subdivision tree. 
                 In the verification step of Newton iterations, use the 
                 $\Tstart$-test of \cite{becker2018near}.
\medskip

In Stage 1, for $c=0,1,\ii$, $\annset_c$ is obtained by computing the 
connected components of the concentric annnuli defined by the output
of $\solveRRC(\pol,c,d^{-2})$.

According to Thm.~\ref{Th_solving_taskS},
Stage 1 involves $
 \app{O}( \deg^2(\deg + \log\|\pol\|_\infty ) )
$ bit operations.

In Stage 2, $\mathcal{D}$ is computed as follows:
using double precision 
floating point arithmetic with correct rounding,
first compute complex discs containing 
all possible intersections of an annulus in 
$\annset_0$ with an annulus in $\annset_1$,
and obtain a set $\mathcal{D}$ of at most $2\deg^2$ complex discs
containing all  roots of $\pol$.
Then, for each disc $\Delta$ in $\mathcal{D}$
 check if $\Delta$ and 
its complex conjugate $\overline{\Delta}$
have a nonempty intersection with at least 
one annulus of $\annset_{\ii}$, and remove $\Delta$ from $\mathcal{D}$
if it does not.
This step has cost in $O(\deg^3)$.

By virtue of 
Prop.~\ref{prop_CzeroCost}, the cost of performing Stage 3
is bounded by the cost of performing 
the algorithm described in Subsec.~\ref{subsec_subdivision}.
This procedure solves the CRC problem
and supports near optimal complexity
for the benchmark problem.

\subsection{Experimental results}
\label{subsec_expeRes_complex}

The procedure of Subsec.~\ref{subsec_algoCRC}
is implemented within \ccluster; below we
call this implementation  \cclusterRR and
present experimental results that highlight
 practical improvement due to using
 our root radii algorithm
in subdivision.

We used \ccluster and \cclusterRR with input value $\varepsilon=2^{-53}$
to find clusters of size at most $\varepsilon$.
We also used \mpsolve{\tt-3.2.1}, with options
{\tt -as -Ga -o16 -j1}
to find approximations with 16 correct digits
of the roots.

For our test polynomials 
(see \ref{subsubsec_testSuite})
we report in Tab.~\ref{table_CRCAlgo}:\\
\noindent - $d$ and $\tau$ denoting  the degree and the bit-size, respectively,\\
\noindent - $t_1$ (resp. $t_2$), the running time of \ccluster (resp. \cclusterRR),\\
\noindent - $n_1$ (resp. $n_2$), the number of $\Tzerot$-tests in \ccluster (resp. \cclusterRR),\\
\noindent - $t_3$, the time
       for computing the three annuli covers in \cclusterRR,\\
\noindent - $t_4$, the running time of \mpsolve in seconds.

For random polynomials, we show averages over 10 examples 
of those values.
We also show $\sigma_1$, $\sigma_2$,
and $\sigma_4$, the standard deviations of the running times
of \ccluster, \cclusterR and \mpsolve.
For the real root isolator presented in Sec.~\ref{section_RRI},
using root radii enables significant  saving of Pellet-based exclusion tests in the subdivision process (compare columns $n_1$ and $n_2$)
and yields a speed-up factor 
about 3  for  our examples (see column $t_2/t_1$).
This speed-up increases as the number of real roots increases
(see, e.g., Wilkinson polynomials)
because some exclusion tests for boxes $B$ 
containing the real line are avoided 
when $2B$ contains at least one root
which we can see from
the number
$n_{\geq 1}(\annset_0,B)$ computed in the $\CCzerot$ test.
The time spent for 
 computing the three annuli covers
remains low compared to the running time of \cclusterR
(see column $t_3/t_2$).
\mpsolve remains the user's choice for 
approximating all complex roots.

\begin{table}[t!]
\centering
\begin{scriptsize}
 \begin{tabular}{rr||cc||ccc||c||c}
       &                & \multicolumn{2}{c||}{\ccluster}  
                        & \multicolumn{4}{c||}{\cclusterRR}
                        & \multicolumn{1}{c}{\mpsolve}\\\hline
   $d$ & $\tau$         & $t_1$ $(\sigma_1)$ & $n_1$
                        & $t_2$ $(\sigma_2)$ & $n_2$ & $t_3/t_2$ (\%)
                        & $t_2/t_1$ (\%)
                        & $t_4$ $(\sigma_4)$ \\\hline
\multicolumn{9}{c}{10 monic random dense polynomials per degree} \\\hline
128 & 128 & 4.43 (.760) & 2598. & \cored{1.46} (.235) & 463. & 7.81 & 33.1  & \coblue{.031} (.003) \\
191 & 191 & 13.5 (1.82) & 3846. & \cored{4.40} (.528) & 694. & 4.20 & 32.6  & \coblue{.063} (.007) \\
256 & 256 & 23.7 (2.52) & 4888. & \cored{7.87} (.672) & 909. & 7.04 & 33.2  & \coblue{.106} (.013) \\
391 & 391 & 70.9 (9.23) & 7494. & \cored{22.5} (1.95) & 1460. & 3.67 & 31.7 & \coblue{.209} (.037) \\
512 & 512 & 154. (17.9) & 9996. & \cored{46.1} (6.00) & 1840. & 7.08 & 29.9 & \coblue{.392} (.102) \\\hline
\multicolumn{9}{c}{Bernoulli polynomials} \\\hline
128 & 410  & 3.86 & 2954  & \cored{1.25} & 548  & 7.48 & 32.3 & \coblue{0.07}\\
191 & 689  & 12.2 & 4026  & \cored{4.51} & 942  & 8.07 & 36.8 & \coblue{0.16}\\
256 & 1056 & 24.7 & 5950  & \cored{10.1} & 1253 & 6.57 & 41.1 & \coblue{0.39}\\
391 & 1809 & 75.1 & 8322  & \cored{27.4} & 1907 & 16.2 & 36.5 & \coblue{0.97}\\
512 & 2590 & 133. & 11738 & \cored{49.9} & 2645 & 12.7 & 37.5 & \coblue{2.32}\\\hline
\multicolumn{9}{c}{Wilkinson polynomials} \\\hline
128 & 721  & 8.43 & 3786  & \cored{1.09} & 14 & 14.4 & 12.9 & \coblue{0.17}\\
191 & 1183 & 25.4 & 5916  & \cored{2.99} & 18 & 27.9 & 11.7 & \coblue{0.51}\\
256 & 1690 & 50.7 & 7500  & \cored{6.34} & 18 & 21.7 & 12.4 & \coblue{1.17}\\
391 & 2815 & 201. & 12780 & \cored{23.1} & 22 & 36.2 & 11.4 & \coblue{4.30}\\
512 & 3882 & 379. & 14994 & \cored{51.3} & 22 & 35.6 & 13.5 & \coblue{9.33}\\\hline
\multicolumn{9}{c}{Polynomials with roots on a regular grid} \\\hline
169 & 369  & 7.37 & 3072  & \cored{1.99} & 592  & 4.03 & 27.1 & \coblue{0.05}\\
289 & 741  & 27.1 & 5864  & \cored{10.2} & 1573 & 3.18 & 37.9 & \coblue{0.13}\\
441 & 1264 & 81.4 & 9976  & \cored{24.4} & 1713 & 4.28 & 29.9 & \coblue{0.56}\\
625 & 1948 & 228. & 15560 & \cored{70.2} & 2508 & 15.0 & 30.7 & \coblue{1.16}\\
841 & 2800 & 493. & 19664 & \cored{169.} & 4294 & 5.75 & 34.2 & \coblue{3.84}\\\hline
\multicolumn{9}{c}{Mignotte polynomials} \\\hline
512 & 256  & 88.8 & 9304 & \cored{28.3} & 1611 & 11.0 & 31.8 & \coblue{0.76}\\
512 & 512  & 88.3 & 9304 & \cored{29.3} & 1570 & 9.20 & 33.1 & \coblue{0.79}\\
512 & 1024 & 101. & 9304 & \cored{32.1} & 1647 & 8.62 & 31.7 & \coblue{0.91}\\
512 & 2048 & 106. & 9304 & \cored{33.4} & 1990 & 7.50 & 31.2 & \coblue{1.12}\\
512 & 4096 & 102. & 9304 & \cored{50.1} & 3593 & 4.88 & 49.0 & \coblue{1.10}\\\hline
 \end{tabular}
 \caption{Runs of \ccluster, \cclusterRR and \mpsolve on our test polynomials.}
\label{table_CRCAlgo}
\end{scriptsize}
\vspace*{-1cm}
\end{table}

\bibliographystyle{splncs04}
\bibliography{references}

\end{document}